\RequirePackage{amsthm}
\documentclass[sn-mathphys-num]{sn-jnl}
\usepackage{braket}

\usepackage{graphicx}%
\usepackage{multirow}%
\usepackage{amsmath,amssymb,amsfonts}%
\usepackage{amsthm}%
\usepackage{mathrsfs}%
\usepackage[title]{appendix}%
\usepackage{xcolor}%
\usepackage{textcomp}%
\usepackage{manyfoot}%
\usepackage{booktabs}%
\usepackage{algorithm}%
\usepackage{algorithmicx}%
\usepackage{algpseudocode}%
\usepackage{listings}%

\theoremstyle{thmstyleone}%
\theoremstyle{thmstyletwo}%

\theoremstyle{thmstylethree}%

\begin{document}

\title[Article Title]{Dynamic protected states in the non-Hermitian system}


\author[1,2]{\fnm{Lei} \sur{Chen}}\email{chenlei@alum.imr.ac.cn}

\author[3]{\fnm{Zhen-Xia} \sur{Niu}}\email{niuzhx@zjnu.edu.cn}

\author*[4]{\fnm{Xingran} \sur{Xu}}\email{thoexxr@hotmail.com}

\affil[1]{School of Information, Hunan University of Humanities, Science and Technology, Loudi 417000, China}

\affil[2]{School of Physics and Electronic Science, Zunyi Normal University, Zunyi 563006, China}
\affil[3]{Department of Physics, Zhejiang Normal University, Jinhua 321004, China}
\affil*[4]{School of Science, Jiangnan University, Wuxi 214122, China}


\abstract{The non-Hermitian skin effect and nonreciprocal behavior are sensitive to the boundary conditions, which are unique features of non-Hermitian systems. The eigenenergies will become complex and all eigenstates are localized at the boundary, which is distinguished from the Hermitian topologies. In this work, we theoretically study the dynamic behavior of the propagation of Gaussian wavepackets inside a non-Hermitian lattice and analyze the self-acceleration process of bulk state or Gaussian wavepackets toward the system's boundary. The initial wavepackets will not only propagate toward the side where the eigenstates are localized, but also their momentum will approach to a specific value where the imaginary parts of energy dispersion are the maximum. In addition, if the wavepackets cover this specific momentum, they will eventually exhibit exponentially increasing amplitudes with time evolution, maintaining the dynamic protected condition for an extended period of time until they approach the boundary. We also take two widely used toy models as examples in one and two dimensions to verify the correspondence of the non-Hermitian skin effect and the dynamic protected state.}

\keywords{skin effect,  photonic lattice, non-Hermitian Hamiltonian,  dynamic behavior}



\maketitle

\section{Introduction}\label{section:one}

For an isolated quantum system, to ensure the conservation
of energy and probability, a Hermitian Hamiltonian is necessary. When
a quantum system couples to its environment, hence, particle loss
and decoherence leads to the breakdown of Hermiticity and requires
non-Hermitian descriptions \cite{Carmichael1993,PhysRevLett.80.5243,Bender1999,Bender2002,Ashida2020}.
It has been shown that non-Hermitian considerations not only provide
suitable descriptions of open systems,  but also bring novel physics
and unprecedented phenomena and applications, as found in a variety
of physical realms including parity-time symmetry and spectral singularity
\cite{Bender2007,El-Ganainy:2018tc,Oezdemir2019,Miri2019}, non-reciprocal and
chiral transport \cite{Fruchart2021}, as well as non-Hermitian topology
and unconventional band theory \cite{Kawabata2019,Zhou2019,Kazuki2019,yao2018}.
Within this context, particularly fascinating phenomenon is the recently
discovered non-Hermitian skin effect (NHSE) \cite{yao2018,Kunst2018,Lee_2019,McDonald2018}.

The NHSE,  an anomalous localization of extensive eigenstates at the open boundaries of a non-Hermitian system, is drastically different from the dynamics of  the extended Bloch waves in Hermitian systems.
This phenomenon was first discovered by Hatano and Nelson \cite{HNmodel1,HNmodel2}
in the late 1900s, where they proposed a 1D disordered tight-binding
lattice model with nearest-neighbour nonreciprocal hopping and showed
that non-Hermiticity induced by the nonreciprocity can prevent Anderson
localization, opening up a mobility region characterized by unidirectional
transportation. Recently, similar phenomena have also been observed
in many seminal works on the topological properties of non-Hermitian
systems. They either revisit the Hatano-Nelson (HN) model to realize
versatile directional transports \cite{Longhi2015,Longhi2015a},
interpret NHSE from different perspectives \cite{MartinezAlvarez2018,skinmode1,Topoorg,Longhi2019,song2019,Yokomizo2021},
even to develop new theories, methods, and material designs to describe
NHSE related topological properties and the associated non-Hermitian
bulk-boundary correspondence (BBC) \cite{Kawabata2019,Kunst2018,Jiangping2020,Xu2021a,Claes2021,Okuma2021},
or extend NHSE to higher dimensions \cite{Yao_2018,fu2020nonhermitian,Kim2021,Ghorashi2021},
to explore new NHSE features enriched by heterogeneous degrees of freedom
\cite{Okuma2019,Xu2021,Lin_2021,Sun2021,Yuce2021},
and experimentally demonstrating the NHSE \cite{zhang2021universal,zhang2021acoustic,Liang2022,Shang2022}.

One of the significant features of the NHSE is that the eigenstates are sensitive to the boundary conditions. The bulk properties of the open boundary condition (OBC) Hamiltonian can be well approximated by the Bloch Hamiltonian with periodic boundary conditions
(PBCs), which is called the BBC in band theory \cite{PhysRevLett.117.076804}. However, the eigenstates with the NHSE cannot be characterized by the BBC because {\it all} of them are localized at the edge with OBCs in the finite-size sysytem \cite{Helbig_2020,Liu2020,Lee2019}. In other words, the eigenstates and eigenenergies with the NHSE under different boundary conditions are completely different from each other. To solve this conflict, the generalized Brillouin zone (GBZ) \cite{Jiangping2020,Xiao_2020,Imura2019} and biorthogonal BBC \cite{Kunst2018,Xu_2020nh} have been proposed. GBZ theory takes the complex momentum and modifies the Bloch factor to obtain the OBC eigenenergies ~\cite{Kunst2018,Kawabata2019,Masahito2018,yao2018,Lieu2018}. Therefore, the corresponding eigenstates are exponentially localized at the edges of the system.  Biorthogonal BBC theory uses both the left and right eigenstates to obtain the OBC eigenenergies from the PBC Hamiltonian. Moreover, the topological invariants of the system with the NHSE are also quite different. The winding number can be defined from the complex energy spectrum, where the complex energy spectrum under PBCs can form a loop and encircle all of the complex energy spectrum under OBCs \cite{Xu2021,XU2022}.

The largest influence of the NHSE on the dynamic properties of the system is reflected in the appearance of a directional (chiral) bulk flow and persistent currents \cite{PhysRevB.105.245143,Longhi2019,skinmode1}. The wavefunctions inside the bulk have a nonreciprocal behavior  where their propagation has a specific direction, which can be detected by experiments \cite{Weidemann_2022,PhysRevLett.129.113601}. In addition, a novel bulk dynamic signature of the NHSE arises in the early stage of the time evolution, i.e., self acceleration of the wavefunction \cite{PhysRevLett.129.113601,Parto_2023}. The complex energy spectrum of the eigenenergies will make the density of the wavepackets exhibit gain or loss for different momentums.
The long time dynamics of the wavefunction are dominated by the interference of Bloch modes with the largest imaginary part of the energy, leading to a directional flow of excitation along the lattice at a constant drift velocity in systems, and displaying the NHSE. However, the imaginary parts of the eigenenergies can have more than one maximum (beyond zero), where the flow of condensates will have a competitive behavior between the maximum points of the eigenenergies.

In this work, we study the dynamic evolution of wavepackets with different velocities in non-Hermitian lattices.
We find that the wavepackets with the maximum imaginary eigenenergies will be protected by the bulk until they reach the boundary. We discuss these dynamic protected states in several models. The paper is organized as follows. In Sec. \ref{Sec2}, we introduce the simplest lattice with the NHSE and discuss the influence of the finite size on the energy spectrum and localization. In Sec. \ref{Sec3}, we investigate the bulk protected states under OBCs and give the analytical result. In Sec. \ref{Sec4}, we take two widely used models  as examples to prove the universality in the non-Hermitian system. In Sec. \ref{Sec5}, we conclude with a summary of our main results and final remarks.

\section{ Finite size and the non-Hermitian skin effect}\label{Sec2}

\begin{figure}[h!]
\centering
\includegraphics[width=\columnwidth]{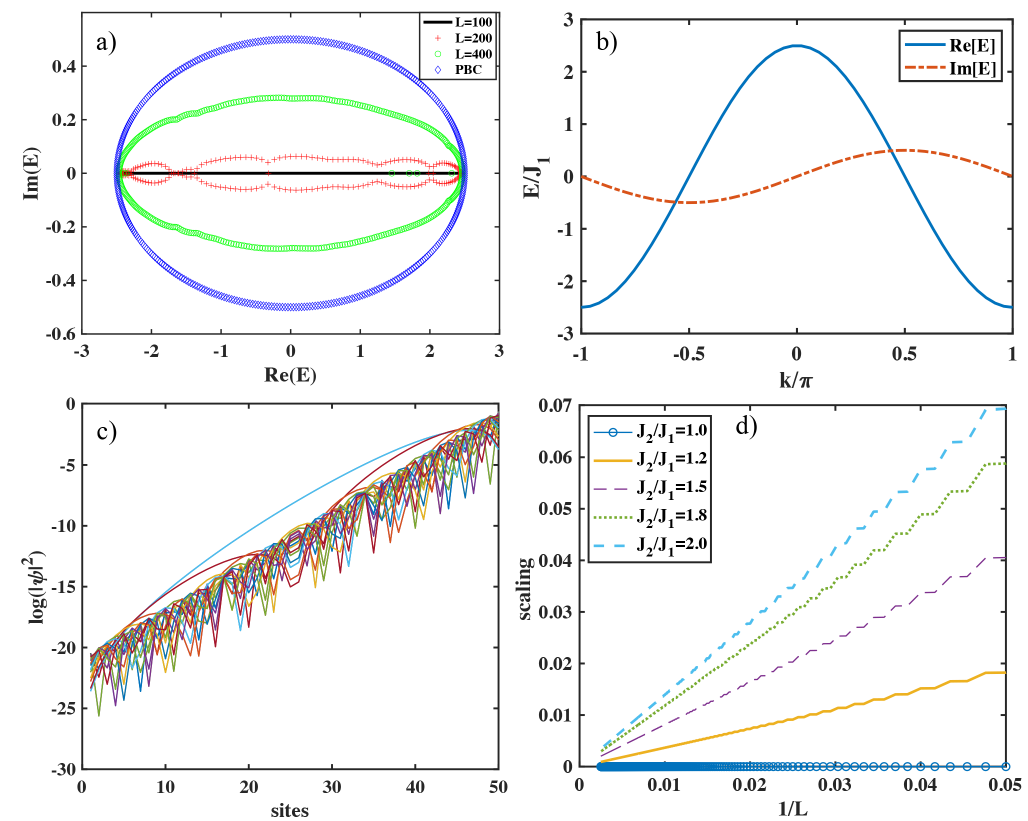}
\caption{a) Complex energy spectrum of the HN model with different system sizes and b) eigenenergies with PBCs.   c) Log function of the density of the eigenstates,  and d) the scaling parameter $\kappa$ as a function of  the system size $L$ in Eq.  (\ref{scaling}).   We take $J_1=1$ and $J_2/J_1=1.5$ for a), b) and c). }\label{HNE}
\end{figure}

In this section, we review the influence of  the size of the system with  NHSE. The HN model is the simplest NHSE model, with
\begin{equation}
H=\sum_n \left( J_1 C^{\dagger}_{n+1}C_n+ J_2 C^{\dagger}_{n}C_{n+1}\right),
\end{equation}
where $C_n(C_n^\dagger)$ is the  annihilation (creation) operator at the $n$  site index,  $J_1$ and $J_2$ are the nearest-neighbor hopping energies along different directions. 

The eigenenergies of the non-Hermitian skin Hamiltonian are always different with different boundary conditions. In particular, the periodic boundary eigenenergies can encircle all open boundary eigenenergies As shown in Fig. \ref{HNE} a), the HN model fundamentally gives purely real eigenenergies when the system has a small size $L$. However, when the system size increases, imaginary parts of the eigenenergies  appear and approach the PBC result (blue signs in Fig. \ref{HNE}a)). Based on this, the topological invariants can easily be defined \cite{Topoorg,Zhang2022review} by
\begin{equation}
W= \frac{1}{2\pi i} \oint_C \frac{d}{dk} \log\left[ Det\left( \hat{H(k)}-E_R \hat{I}  \right)\right], \label{winding}
\end{equation}
where $H(k)=\sum_k \left[ (J_1+J_2)\cos k-i(J_1-J_2)\sin k\right] C_k^\dagger C_k$,  can be obtained by the Fourier transformation of the real-space Hamiltonian (see Fig. \ref{HNE}b) ) and $E_R$ is the reference energy. The eigenenergies are complex when the system size is large enough or the system has PBCs.  The winding number $W$ is dependent on  the sign of $J_1-J_2$.  If $J_1=J_2$, the winding number is zero, all eigenenergies are real,  and the NHSE vanishes. However, $J_1\neq J_2$ will let the topological  invariants arise and all eigenstates localize at the  side. 

In the HN model, all wavefunctions are localized at the edge. The amplitudes of the absolute value of the density will exponentially increase or descrease (depending on the magnitude of $J_1$ and $J_2$) with the site index $n$, as  illustrated in Fig. \ref{HNE} (c).  The relationship between the density of the first site and the last site can be described by 
\begin{equation}
|\psi(n)|^2=e^{\kappa n}|\psi(1)|^2 \label{scaling},
\end{equation} 
where $\kappa$ is the scaling parameter and $n$ is the site index.  In a Hermitian lattice, the Bloch factor $|e^{ik}|=1$, however, GBZ theory states that $|e^{ik}|=\beta\neq1$, where all wavefunctions can have exponential localization along the index.  So the translation symmetry is broken in the non-Hermitian lattice \cite{scaling2023,scalingfang} .  As is illustrated in Fig. \ref{HNE} (d), the scaling paramters $\kappa$ is proportional to the inverse of the system size $\kappa \sim L^{-1}$.  Therefore, the localization of the density will vanish when the system size goes to infinity.

\section{Dynamic protected state}\label{Sec3}

In this section, we will analyze the time evolution of the Gaussian wavepackets and obtain the motion of the center mass in the initial stage of evolution.
\begin{figure}
\centering
\includegraphics[width=\columnwidth]{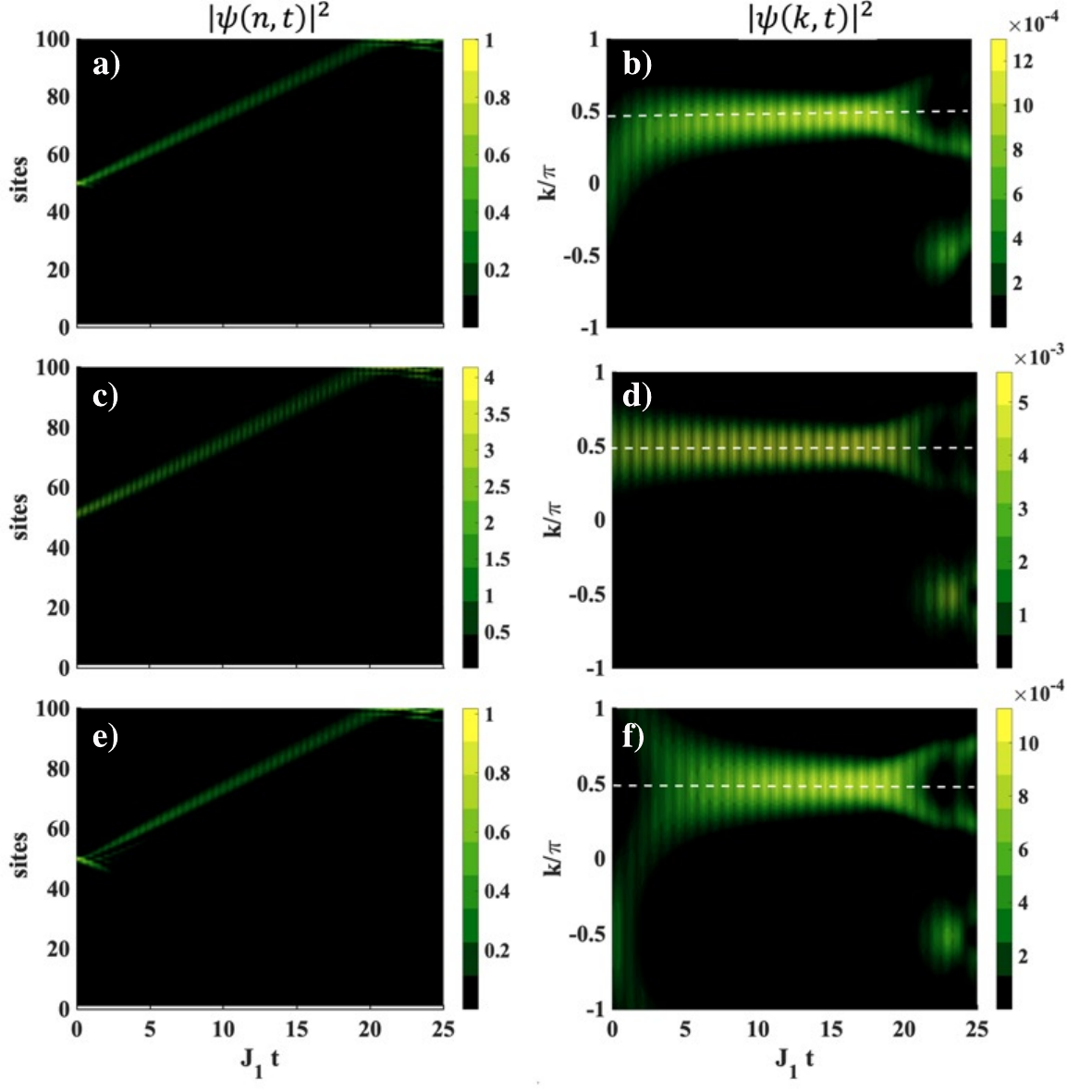}
\caption{Time evolution of the Gaussian wavepackets with different initial momenta in real space (first column) and momentum space (second column). The parameters are $J_2/J_1$=$1.5$, $k_0$=$0$, and $-\pi/2$ or $\pi/2$ for different rows.}\label{WF}
\end{figure}

In a Hermitian topological system, only the edge states are topologically protected, and their propagation is chiral. However, all eigenstates of the non-Hermitian Hamiltonian are localized, including the bulk and edge states.  The wavefunction is forbidden to propagate to one side and get enhanced to propagate to the opposite side in the 1D case\cite{PhysRevB.108.115404,Malzard_2018}.

When we consider the time evolution of the wavepackets of the system, we need to expand the initial state with the eigenstates. The eigenenergies are complex for a non-Hermitian system with PBCs or a sufficiently large size. The wavepacket with momentum $k$ evolutes with time as follows,
\begin{equation}
\ket{\psi(k,t)}=e^{-i {Re} (E) t}e^{ {Im}(E) t}\ket{\psi(k,0)},
\end{equation}
where the real eigenenergies (${Re} (E)$) govern the motion of the wavepacket and the imaginary eigenenergies describe the decay or growth rate of the corresponding eigenstates. When ${Im}(E)<0$ or ${Im}(E)>0$, the corresponding eigenstate will exponentially decay or grow with time. Therefore, the time evolution of the wavefunction will have a peculiar dynamic behavior where all wavefunctions in momentum space will become the same.

The time evolution of the wavefunction in momentum space can be obtained by the Fourier transformation, with
\begin{equation}
\psi(k,0)=\sum_{n=0}^{L-1}  \psi(n,0) e^{-i k n}.
\end{equation}
As shown in Fig. \ref{WF}, we place the Gaussian state with different initial velocities $k_0$  in the center of the HN lattice . The wavepacket has nonreciprocal behavior, and its center of mass only moves to one side. In addition, if the initial momentum is not equal to the momentum $k_m=\pi/2$ where the imaginary eigenenergies have the maximum, then the momentum of the wavepacket will approach $k_m$. Specifically, when we use $k_0=k_m$ as the momentum of the initial state, the Gaussian state is protected until it reaches the boundary which is unique compared to  the Hermitian topological lattice. Due to the maximum of the imaginary energies in the non-Hermitian lattice with PBCs, only the wavefunction with momentum $k_m$ will be the most enhanced along with the time evolution.  Therefore, the propagation of the wavepacket with OBCs in the bulk can still obey the rule of PBCs and survive for a sufficiently long time until it reaches the boundary.

\begin{figure}[h!]
\centering
\includegraphics[width=\columnwidth]{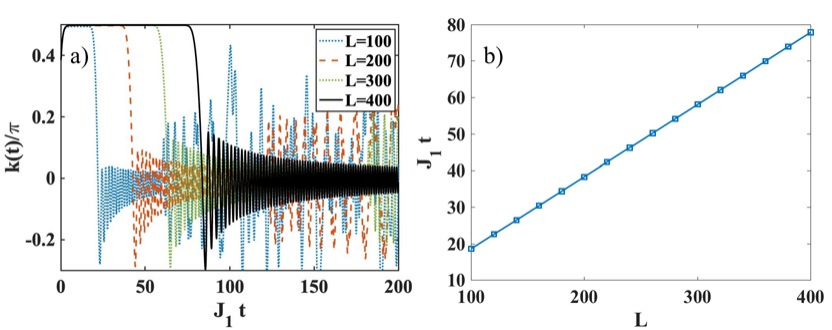}
\caption{The mean momentums of the wavepackets  changes with time  at different system sizes $L$ a),  and  the duration of the dynamic protected state as a function of the  system size b). Parameters are $J_2/J_1=1.5$ and $k_0=0.4\pi$. }\label{HNDU}
\end{figure}

The duration of the wavepacket with OBCs linearly increases with the system size, as illustrated in Fig. \ref{HNDU}. When we set the momentum of the initial state near $k_m$ ($k_0=0.4\pi$ ),  the speed of wavepacket $k(t)$ will keep a duration at $k_m$, and then maintain a long time unless reaching the boundary. The initial state in momentum space can be described as
\begin{equation}
\psi(k,0)=e^{(k-k_0)^2/2\sigma^2},
\end{equation}
where $k_0$ is the center of the wavepacket in $k$ space and $\sigma $ is the width. When the Gaussian state is in the bulk, we can use the PBC Hamiltonian to explain the wavepacket dynamic properties , with
\begin{equation}
\psi(k,t)=e^{-it \sum_k (J_1+J_2)\cos k-(J_1-J_2)\sin k} e^{(k-k_0)^2/2\sigma^2}.
\end{equation}
The first term is the oscillation frequency and can be ignored compared to the exponential decay or gain due to the imaginary eigenenergies.

We consider a non-Hermitian Hamiltonian, denoted as $\hat{H} = \sum_k E_k \ket{k}\bra{k}$, where $\ket{k}$ represents the right eigenvector with  eigenenergy $E_k$.
An initial state $\ket{\psi(t=0)}$ is governed by the evolution operator $U(t)$ associated with this Hamiltonian. The state at time $t$, denoted as $\ket{\psi(k,t)}$, is determined by the evolution operator $U(t)$ associated with the Hamiltonian. Mathematically, it can be expressed as
\begin{eqnarray}
\ket{\psi(k,t)}& =& \hat{U}(t) \ket{\psi(k,0)} = e^{-i \hat{H} t}\ket{\psi(k,0)} \nonumber \\
&=& \prod_k e^{-i E_k t \ket{k}\bra{k} }\ket{\psi(k,0)}.
\end{eqnarray}
Let us examine the variation in the mean momentum over time, given by $\bra{\psi(k,t)} \hat{k} \ket{\psi(k,t)}$.
\begin{eqnarray}\nonumber
k(t) &=& \bra{\psi(k,t)} \hat{k} \ket{\psi(k,t)}\\\nonumber
&=& \bra{\psi(k,0)} \prod_{k_2} e^{i E^*_{k_2} t \ket{k_2}\bra{k_2} } \hat{k} \prod_{k_1} e^{-i E_{k_1} t \ket{k_1}\bra{k_1} } \ket{\psi(k,0)} \\\nonumber
&=& \int dk' \braket{\psi(k,0)|k'} e^{i (E^*_{k'} -E_{k'}) t } k' \braket{k'|\psi(k,0)} \\
&=& \int dk' e^{ 2 {Im} E_{k'}t } k' |\braket{k'|\psi(k,0)}|^2,
\end{eqnarray}
where we insert a unit operator $\int dk' \ket{k'} \bra{k'}$. In Hermitian systems, $E_{k'}$ is purely real, observing that $k(t) = k(0)$ remains unchanged with time evolution.
In non-Hermitian systems, when ${Im}(E_{k'}) > 0$, the contribution of $k'$ becomes more significant,  while $ {Im}(E_{k'}) < 0$, the contribution of $k'$ diminishes.
A predictable consequence is that $k(t)$ eventually approaches $k_m$, which corresponds to the maximum value of ${Im}(E_{k_m})$.
Without loss of generality, we can shift ${Im}(E_k)$ such that the maximum point corresponds to zero. We can obtain
\begin{equation}
k(t) = \int dk' e^{ 2{Im} (\tilde{E}_{k'})t } e^{ 2at } k' |\braket{k'|\psi(k,0)}|^2,
\end{equation}
where ${Im} (\tilde{E}_{k'}) = {Im} (E_{k'} - a)$, where $a$ is a positive constant. We consider a simplified case where $a = 0$, indicating that only dissipative effects are present in the system.
The contribution of all values of $k$ except for the maximum point decreases. Therefore, as long as $|\braket{k_m|\psi(0)}|^2$ is nonzero, $\lim_{t\to \infty} k(t)$ remains equal to $k_m$.

Therefore, the dynamic protected states are the wavepacket that will not change with time during the propagation period.  The momentum of the wavepackets will approach to the $k_m$, which is the maximum point of the  imaginary parts of the eigenenergies of the non-Hermitian Hamiltonian with NHSE. 

\section{Dynamic protected states of different non-Hermitian systems}\label{Sec4}

In this section, we take two widely used theoretical models (the SSH model and 2D polariton system) with different symmetries as examples to check the universality of the dynamic protected state in non-Hermitian systems and discuss the influence of the skin effect. The time and energy dimensions are normalized with $J_1t$ and $E/J_1$.

\subsection{SSH model with chiral symmetry}
\begin{figure}[h!]
\centering
\includegraphics[width=0.8\columnwidth]{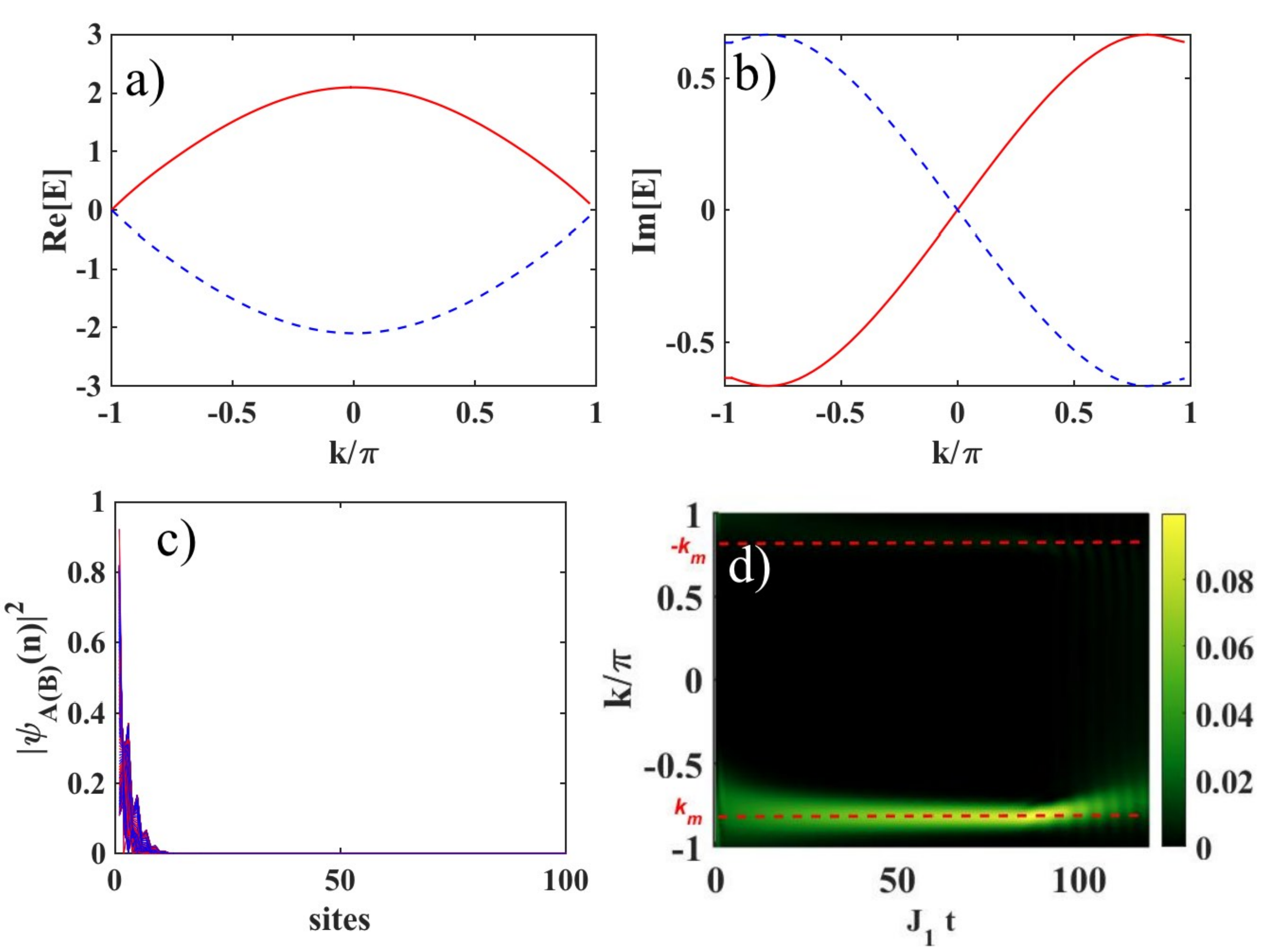}
\caption{  The real a) and the imaginary b) parts of the eigenenergies of the SSH model with different spins. Eigenfunctions of different spins c), and the density evolution of spin-up with the initial velocity $k_0=-0.5\pi$ d).  The other parameters are $t_2/t_1=1.2$ and $\gamma/t_1=4/3$. }\label{SSH}
\end{figure}
The SSH model with different nearest-neighbor hopping energies in a 1D lattice is widely studied to realize the NHSE \cite{ Liu_2020,Fu_2020}. The Hamiltonian in momentum $k$ space can be described as
\begin{equation}
H=d_x \sigma_x+(d_y+i\frac{\gamma}{2})\sigma_y, \label{ssheq}
\end{equation}
where $\gamma/2$ changes the hopping term in the unit cell with different hopping strengths.   $d_x=t_1+t_2\cos k$, $d_y=t_2 \sin k$ and $\sigma_{x,y}$ are the Pauli matrices. $t_1$ and $t_2$ are the hopping between the different spins in the same unit cell and the hopping between the nearest unit cells. 

The real parts and imaginary parts of eigenvalues of  Eq. (\ref{ssheq}) with the Fourier transformation in momentum space are shown in Figs.  \ref{SSH} a)-b).   All eigenstates with OBCs are localized at the left side  (shown in Fig.  \ref{SSH} c) ) when $t_1<t_2$. The eigenvalues appear in pairs, with
\begin{equation}
E_{\pm}(k)=\pm\sqrt{ t_1^2+t_2^2+2t_1t_2\cos k-\gamma^2/4+it_2\gamma\sin k}.
\end{equation}
The OBC eigenenergies are zero modes in the finite-size system,  and GBZ theory gives the non-Bloch factor $|e^{ik}|= \sqrt{ \left| \frac{t_1-\gamma/2}{t_1+\gamma/2}\right|} \neq 1$, where $k$ becomes a complex number and makes the eigenstates localized.

The imaginary parts of eigenvalues of different spins have different maximum points $k_m$ and $-k_m$ (red solid line and blue dashed line in  Fig. \ref{SSH} b)) with PBC.    When we place the initial Gaussian state in the middle of the lattice with OBC,  the mean momentum of the wavepacket will  have two peaks $k_m$ and $-k_m$.  However,  the density of $| \psi(k,t)|^2$ are most localized at $k_m$ rather than $-k_m$ as shown in Fig. \ref{SSH} d).   The above results imply that the  boundary condition plays a significant role in the non-Hermitian system with skin effect.  The eigenstates of the Hamiltonian with OBC are localized at the left side (shown in Fig.  \ref{SSH} c) ),  so  dynamic behaviors of a condensation in the bulk are also affected by the boundary condition and propagate to the left side with momentum $k_m$.

\subsection{ Two-dimensional non-Hermitian system}

\begin{figure}[h]
\centering
\includegraphics[width=0.8\textwidth]{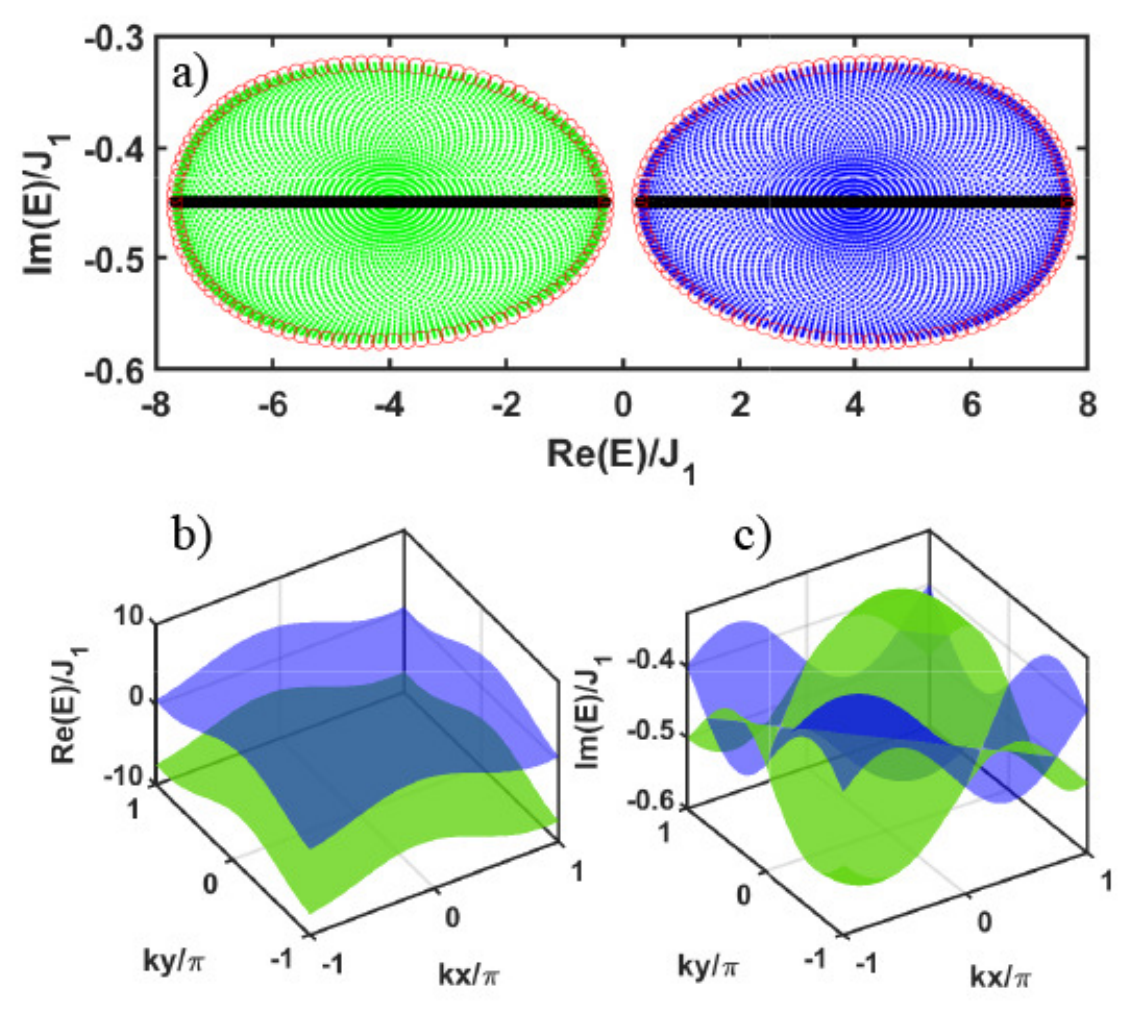}\\
\caption{Complex energy spectrum with PBCs and OBCs a), and real b) and imaginary c) energy bands of PBC eigenenergies. The black lines are the OBC energies, and the blue and green areas are filled with the PBC energies of the upper and lower bands. Meanwhile, the green dots are the complex energies along with the selected integration direction for defining the winding number in (a). Parameters: $J_2/J_1$=1, $k_p =\pi/4$, $\gamma_+/J_1$=$0.1$, $\gamma_-/J_1$=$0.8$, and $A/J_1$=$4$. }\label{cornerES}   
\end{figure}

\begin{figure*}
\centering
\includegraphics[width=1.0\textwidth]{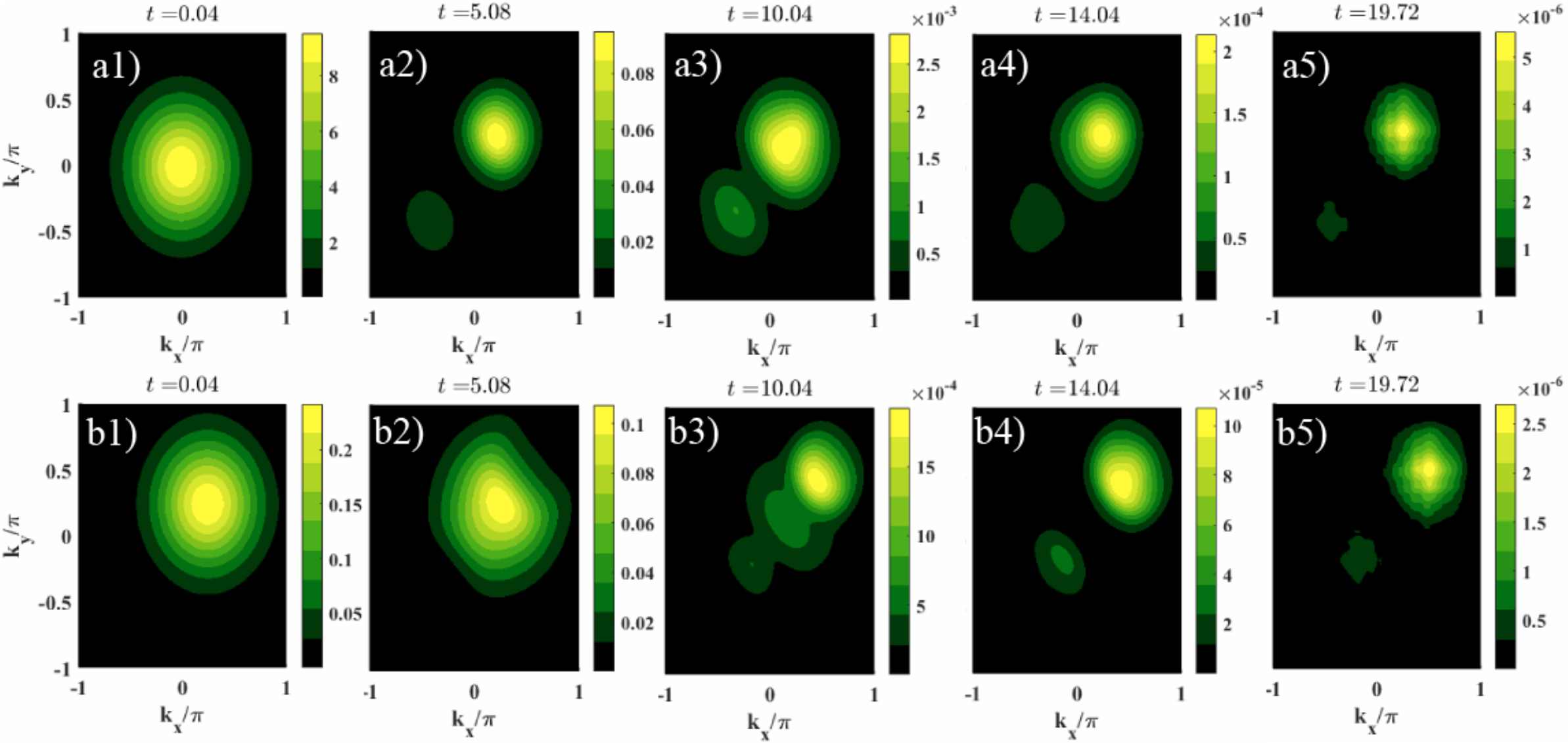}\\
\caption{Time evolution of the Gaussian states in momentum space for different spins (different rows) obtained with Hamiltonian Eq. (\ref{cornereq}). Parameters: $J_2/J_1$=1, $k_p =\pi/4$, $\gamma_+/J_1$=$0.1$, $\gamma_-/J_1$=$0.8$, and $A/J_1$=$4$. }\label{cornertime}
\end{figure*}

In this subsection we will discuss dynamic behaviors and preotected states of the higher-order NHSE in 2D polariton system.  The previous work \cite{XU2022} have revealed that all wavefunctions are localized at the corners in 2D tight-binding polariton model.  The effective Hamiltonian can  be described as
\begin{eqnarray}
H_{m,n}&=&\sum_{m,n}\left[J_{1}\left(\hat{a}_{m,n}^{\dagger}\hat{a}_{m,n+1}+\hat{b}_{m,n}^{\dagger}\hat{b}_{m,n+1}+h.c.\right)+J_{2}\left(\hat{a}_{m+1,n}^{\dagger}\hat{a}_{m,n}+\hat{b}_{m+1,n}^{\dagger}\hat{b}_{m,n}+h.c.\right)\right] \nonumber\\
&+&\sum_{m,n}\left[-i\gamma_{+}\left(\hat{a}_{m,n}^{\dagger}\hat{a}_{m,n}\right)-i\gamma_{-}\left(\hat{b}_{m,n}^{\dagger}\hat{b}_{m,n}\right)\right] \nonumber \\
&+&\sum_{m,n}\left[A\hat{a}_{m,n}^{\dagger}\hat{b}_{m,n}e^{-ik_{p}\left(m+n\right)}+A\hat{b}_{m,n}^{\dagger}\hat{a}_{m,n}e^{ik_{p}\left(m+n\right)}\right],  \label{cornereq}
\end{eqnarray}
where $\hat{a}_{m,n}$ ($\hat{a}^\dagger_{m,n}$) and $\hat{b}_{m,n}$ ($\hat{b}^\dagger_{m,n}$) denote the annihilation (creation )of particles with spin $\pm$ at site $(m,n)$.  $J_{1,2}$ are the hopping energies in the $x$ and $y$ directions, $\gamma_{\pm}$ are the different decay (or gain) rates of different spins, and $A$ describes the strength of the polarization  splitting.

As shown in Fig. \ref{cornerES} a),  the eigenenergies with OBC (black dots) are  surrounded by the PBC energies (green and blue areas).  All wavefunctions are localized at the off-side corners  and the winding number is dependent on the selected paths.  Furthermore,  only along with the path  $k_x=k_y$ (red circles),  all OBC eigenenergies are encircled by the PBC eigenenergies in this path \cite{XU2022}.     The eigenenergies of different spins are plotted with different colors (green and blue areas) and they are separated from each other.  The winding number for each spin is opposite and their corresponding eigenstates are localized at the opposite corners \cite{XU2022}.  The system can be transformed into a two-band model with PBC and the real and imaginary parts of the corresponding eigenenergies as shown in Figs. \ref{cornerES} b) and c). The real parts are separated from each other, but the imaginary parts have crossovers.   The higer order NHSE is different from the 1D case since the winding number is determined by the chosen integrated path. 

The imaginary parts of the eigenenergies of Eq. (\ref{cornereq}) have multiple maximum points as illustrated in Fig. \ref{cornerES} c).    To check the density distrubution in the momentum space,  we place two Gaussian states  with zero initial speeds at the middle of the lattice  in Fig. \ref{cornertime}. The density distribution of two spins will gain velocities  because of the self-accelaration effect \cite{PhysRevB.105.245143}.  However,  all eigenstates of the OBC Hamiltonian are localized at the off-side corners.   Here, the effective decays for different spins $\gamma_\pm$ have the crucial effect for letting the density decay along with different directions are different in the time evolution.

At the begining, the density distribution for different spins in momentum space are localized at the different $k_x$ and $k_y$ as shown in Figs. \ref{cornertime} a1)-(a2) and b1)-b2).  After the sufficient long time,  the two-dimensional Gaussian states in momentum space  will remain stable as shown in Figs.  \ref{cornertime} a3)-a5) and b3)-b5).   In addition, the density distributions for different spins are localized at the same $k_x$ and $k_y$ although the imaginary parts of the eigenenergies of the PBC Hamiltonian are different and have different maximum points.   The eigenstates of the non-Hermitian system are sensitive to the boundary condition  and so do the dynamic protected states.  

\section{Discussion}\label{Sec5}
In this work, we study the time evolution of the wavepackets in a non-Hermitian lattice. The initial states are prepared as Gaussian states and  have different center-mass positions and initial velocities. The  mean momentum of Gaussian wavepacket will approach to the certain momentum, at which the eigenenergies have the largest imaginary parts.  This dynamic protected state is stable  in the bulk with OBC and will exponentially accelerate from the initial speed.  Furthermore, this state will keep the certain momentum for a long time during the time evolution until the wavepackets reach the boundary. We also take the 1D SSH and 2D polariton system models  as examples whose eigenenergies have more than one maximum points. The momentum of Gaussian wavepackets will only have one momentum depending on which sides are the eigenstates localized and also sensitive to the boundary condition.

On the one hand, our results provide a better understanding for the non-Hermitian skin effect and self acceleration of the wavefunction. On the other hand, the sensitivity and robustness of the evolution of bulk states with different momentum in the non-Hermitian lattices are of great significant to design new quantum devices, such as making the electronic sensor with a high-level sensitivity and a strong robustness \cite{Yuan_2023,Zhang_2024}, advancing non-Hermitian topological magnonics and photonics to deal with the engineering of dissipation and/or gain for non-Hermitian topological phases \cite{Yu_2024,Yan_2023, Li_2023nn} , among other \cite{Zhao_2019,Gu_2022}.

\section{Acknowledgements}\label{Sec6}
This work is supported by the Fundamental Research Funds for the Central Universities ( Grant No. JUSRP123027 ), National Natural Science Foundation of China ( Grant No. 12264061 ), the Zhejiang Provincial Nature Science Foundation under ( Grant No.~LQ22A040006 ), and the Science Foundation of Guizhou Science and Technology Department ( Grant number QKHJZ[2021]033 ).

\section{Competing Interests}
The authors declare no competing interests.

\section{Author Contribution Statement}
Lei Chen and Zhen-Xia Niu carried out the analytical calculation. Xingran Xu did the numerical simulation and supervised the project.

\section{Data Availability Statement}
The datasets used and analysed during the current study are available from the corresponding author on reasonable request.


\bibliography{mybib}


\begin{thebibliography}{72}
\ifx \bisbn   \undefined \def \bisbn  #1{ISBN #1}\fi
\ifx \binits  \undefined \def \binits#1{#1}\fi
\ifx \bauthor  \undefined \def \bauthor#1{#1}\fi
\ifx \batitle  \undefined \def \batitle#1{#1}\fi
\ifx \bjtitle  \undefined \def \bjtitle#1{#1}\fi
\ifx \bvolume  \undefined \def \bvolume#1{\textbf{#1}}\fi
\ifx \byear  \undefined \def \byear#1{#1}\fi
\ifx \bissue  \undefined \def \bissue#1{#1}\fi
\ifx \bfpage  \undefined \def \bfpage#1{#1}\fi
\ifx \blpage  \undefined \def \blpage #1{#1}\fi
\ifx \burl  \undefined \def \burl#1{\textsf{#1}}\fi
\ifx \doiurl  \undefined \def \doiurl#1{\url{https://doi.org/#1}}\fi
\ifx \betal  \undefined \def \betal{\textit{et al.}}\fi
\ifx \binstitute  \undefined \def \binstitute#1{#1}\fi
\ifx \binstitutionaled  \undefined \def \binstitutionaled#1{#1}\fi
\ifx \bctitle  \undefined \def \bctitle#1{#1}\fi
\ifx \beditor  \undefined \def \beditor#1{#1}\fi
\ifx \bpublisher  \undefined \def \bpublisher#1{#1}\fi
\ifx \bbtitle  \undefined \def \bbtitle#1{#1}\fi
\ifx \bedition  \undefined \def \bedition#1{#1}\fi
\ifx \bseriesno  \undefined \def \bseriesno#1{#1}\fi
\ifx \blocation  \undefined \def \blocation#1{#1}\fi
\ifx \bsertitle  \undefined \def \bsertitle#1{#1}\fi
\ifx \bsnm \undefined \def \bsnm#1{#1}\fi
\ifx \bsuffix \undefined \def \bsuffix#1{#1}\fi
\ifx \bparticle \undefined \def \bparticle#1{#1}\fi
\ifx \barticle \undefined \def \barticle#1{#1}\fi
\bibcommenthead
\ifx \bconfdate \undefined \def \bconfdate #1{#1}\fi
\ifx \botherref \undefined \def \botherref #1{#1}\fi
\ifx \url \undefined \def \url#1{\textsf{#1}}\fi
\ifx \bchapter \undefined \def \bchapter#1{#1}\fi
\ifx \bbook \undefined \def \bbook#1{#1}\fi
\ifx \bcomment \undefined \def \bcomment#1{#1}\fi
\ifx \oauthor \undefined \def \oauthor#1{#1}\fi
\ifx \citeauthoryear \undefined \def \citeauthoryear#1{#1}\fi
\ifx \endbibitem  \undefined \def \endbibitem {}\fi
\ifx \bconflocation  \undefined \def \bconflocation#1{#1}\fi
\ifx \arxivurl  \undefined \def \arxivurl#1{\textsf{#1}}\fi
\csname PreBibitemsHook\endcsname

\bibitem[\protect\citeauthoryear{Carmichael}{1993}]{Carmichael1993}
\begin{barticle}
\bauthor{\bsnm{Carmichael}, \binits{H.J.}}:
\batitle{Quantum trajectory theory for cascaded open systems}.
\bjtitle{Phys. Rev. Lett.}
\bvolume{70}(\bissue{15}),
\bfpage{2273}
(\byear{1993})
\doiurl{10.1103/physrevlett.70.2273}
\end{barticle}
\endbibitem

\bibitem[\protect\citeauthoryear{Bender and
  Boettcher}{1998}]{PhysRevLett.80.5243}
\begin{barticle}
\bauthor{\bsnm{Bender}, \binits{C.M.}},
\bauthor{\bsnm{Boettcher}, \binits{S.}}:
\batitle{Real spectra in non-hermitian hamiltonians having pt symmetry}.
\bjtitle{Phys. Rev. Lett.}
\bvolume{80},
\bfpage{5243}--\blpage{5246}
(\byear{1998})
\doiurl{10.1103/PhysRevLett.80.5243}
\end{barticle}
\endbibitem

\bibitem[\protect\citeauthoryear{Bender et~al.}{1999}]{Bender1999}
\begin{barticle}
\bauthor{\bsnm{Bender}, \binits{C.M.}},
\bauthor{\bsnm{Boettcher}, \binits{S.}},
\bauthor{\bsnm{Meisinger}, \binits{P.N.}}:
\batitle{Pt-symmetric quantum mechanics}.
\bjtitle{J. Math. Phys.}
\bvolume{40}(\bissue{5}),
\bfpage{2201}--\blpage{2229}
(\byear{1999})
\doiurl{10.1063/1.532860}
\end{barticle}
\endbibitem

\bibitem[\protect\citeauthoryear{Bender et~al.}{2002}]{Bender2002}
\begin{barticle}
\bauthor{\bsnm{Bender}, \binits{C.M.}},
\bauthor{\bsnm{Brody}, \binits{D.C.}},
\bauthor{\bsnm{Jones}, \binits{H.F.}}:
\batitle{Complex extension of quantum mechanics}.
\bjtitle{Phys. Rev. Lett.}
\bvolume{89}(\bissue{27}),
\bfpage{270401}
(\byear{2002})
\doiurl{10.1103/physrevlett.89.270401}
\end{barticle}
\endbibitem

\bibitem[\protect\citeauthoryear{Ashida et~al.}{2020}]{Ashida2020}
\begin{barticle}
\bauthor{\bsnm{Ashida}, \binits{Y.}},
\bauthor{\bsnm{Gong}, \binits{Z.}},
\bauthor{\bsnm{Ueda}, \binits{M.}}:
\batitle{Non-hermitian physics}.
\bjtitle{Adv. Phys.}
\bvolume{69}(\bissue{3}),
\bfpage{249}--\blpage{435}
(\byear{2020})
\doiurl{10.1080/00018732.2021.1876991}
\end{barticle}
\endbibitem

\bibitem[\protect\citeauthoryear{Bender}{2007}]{Bender2007}
\begin{barticle}
\bauthor{\bsnm{Bender}, \binits{C.M.}}:
\batitle{Making sense of non-hermitian hamiltonians}.
\bjtitle{Rep. Prog. Phys.}
\bvolume{70}(\bissue{6}),
\bfpage{947}
(\byear{2007})
\doiurl{10.1088/0034-4885/70/6/r03}
\end{barticle}
\endbibitem

\bibitem[\protect\citeauthoryear{El-Ganainy et~al.}{2018}]{El-Ganainy:2018tc}
\begin{barticle}
\bauthor{\bsnm{El-Ganainy}, \binits{R.}},
\bauthor{\bsnm{Makris}, \binits{K.G.}},
\bauthor{\bsnm{Khajavikhan}, \binits{M.}},
\bauthor{\bsnm{Musslimani}, \binits{Z.H.}},
\bauthor{\bsnm{Rotter}, \binits{S.}},
\bauthor{\bsnm{Christodoulides}, \binits{D.N.}}:
\batitle{Non-hermitian physics and pt symmetry}.
\bjtitle{Nature Physics}
\bvolume{14}(\bissue{1}),
\bfpage{11}--\blpage{19}
(\byear{2018})
\doiurl{10.1038/nphys4323}
\end{barticle}
\endbibitem

\bibitem[\protect\citeauthoryear{{\"O}zdemir et~al.}{2019}]{Oezdemir2019}
\begin{barticle}
\bauthor{\bsnm{{\"O}zdemir}, \binits{{\c{S}}.K.}},
\bauthor{\bsnm{Rotter}, \binits{S.}},
\bauthor{\bsnm{Nori}, \binits{F.}},
\bauthor{\bsnm{Yang}, \binits{L.}}:
\batitle{Parity--time symmetry and exceptional points in photonics}.
\bjtitle{Nature materials}
\bvolume{18}(\bissue{8}),
\bfpage{783}--\blpage{798}
(\byear{2019})
\doiurl{10.1038/s41563-019-0304-9}
\end{barticle}
\endbibitem

\bibitem[\protect\citeauthoryear{Miri and Alu}{2019}]{Miri2019}
\begin{barticle}
\bauthor{\bsnm{Miri}, \binits{M.-A.}},
\bauthor{\bsnm{Alu}, \binits{A.}}:
\batitle{Exceptional points in optics and photonics}.
\bjtitle{Science}
\bvolume{363}(\bissue{6422}),
\bfpage{7709}
(\byear{2019})
\doiurl{10.1126/science.aar7709}
\end{barticle}
\endbibitem

\bibitem[\protect\citeauthoryear{Fruchart et~al.}{2021}]{Fruchart2021}
\begin{barticle}
\bauthor{\bsnm{Fruchart}, \binits{M.}},
\bauthor{\bsnm{Hanai}, \binits{R.}},
\bauthor{\bsnm{Littlewood}, \binits{P.B.}},
\bauthor{\bsnm{Vitelli}, \binits{V.}}:
\batitle{Non-reciprocal phase transitions}.
\bjtitle{Nature}
\bvolume{592}(\bissue{7854}),
\bfpage{363}--\blpage{369}
(\byear{2021})
\doiurl{10.1038/s41586-021-03375-9}
\end{barticle}
\endbibitem

\bibitem[\protect\citeauthoryear{Kawabata et~al.}{2019}]{Kawabata2019}
\begin{barticle}
\bauthor{\bsnm{Kawabata}, \binits{K.}},
\bauthor{\bsnm{Shiozaki}, \binits{K.}},
\bauthor{\bsnm{Ueda}, \binits{M.}},
\bauthor{\bsnm{Sato}, \binits{M.}}:
\batitle{Symmetry and topology in non-hermitian physics}.
\bjtitle{Phys. Rev. X}
\bvolume{9},
\bfpage{041015}
(\byear{2019})
\doiurl{10.1103/PhysRevX.9.041015}
\end{barticle}
\endbibitem

\bibitem[\protect\citeauthoryear{Zhou and Lee}{2019}]{Zhou2019}
\begin{barticle}
\bauthor{\bsnm{Zhou}, \binits{H.}},
\bauthor{\bsnm{Lee}, \binits{J.Y.}}:
\batitle{Periodic table for topological bands with non-hermitian symmetries}.
\bjtitle{Phys. Rev. B}
\bvolume{99},
\bfpage{235112}
(\byear{2019})
\doiurl{10.1103/PhysRevB.99.235112}
\end{barticle}
\endbibitem

\bibitem[\protect\citeauthoryear{Yokomizo and Murakami}{2019}]{Kazuki2019}
\begin{barticle}
\bauthor{\bsnm{Yokomizo}, \binits{K.}},
\bauthor{\bsnm{Murakami}, \binits{S.}}:
\batitle{Non-bloch band theory of non-hermitian systems}.
\bjtitle{Phys. Rev. Lett.}
\bvolume{123},
\bfpage{066404}
(\byear{2019})
\doiurl{10.1103/PhysRevLett.123.066404}
\end{barticle}
\endbibitem

\bibitem[\protect\citeauthoryear{Yao and Wang}{2018}]{yao2018}
\begin{barticle}
\bauthor{\bsnm{Yao}, \binits{S.}},
\bauthor{\bsnm{Wang}, \binits{Z.}}:
\batitle{Edge states and topological invariants of non-hermitian systems}.
\bjtitle{Phys. Rev. Lett.}
\bvolume{121},
\bfpage{086803}
(\byear{2018})
\doiurl{10.1103/PhysRevLett.121.086803}
\end{barticle}
\endbibitem

\bibitem[\protect\citeauthoryear{Kunst et~al.}{2018}]{Kunst2018}
\begin{barticle}
\bauthor{\bsnm{Kunst}, \binits{F.K.}},
\bauthor{\bsnm{Edvardsson}, \binits{E.}},
\bauthor{\bsnm{Budich}, \binits{J.C.}},
\bauthor{\bsnm{Bergholtz}, \binits{E.J.}}:
\batitle{Biorthogonal bulk-boundary correspondence in non-hermitian systems}.
\bjtitle{Phys. Rev. Lett.}
\bvolume{121},
\bfpage{026808}
(\byear{2018})
\doiurl{10.1103/PhysRevLett.121.026808}
\end{barticle}
\endbibitem

\bibitem[\protect\citeauthoryear{Lee and Thomale}{2019}]{Lee_2019}
\begin{barticle}
\bauthor{\bsnm{Lee}, \binits{C.H.}},
\bauthor{\bsnm{Thomale}, \binits{R.}}:
\batitle{Anatomy of skin modes and topology in non-hermitian systems}.
\bjtitle{Phys. Rev. B}
\bvolume{99},
\bfpage{201103}
(\byear{2019})
\doiurl{10.1103/PhysRevB.99.201103}
\end{barticle}
\endbibitem

\bibitem[\protect\citeauthoryear{McDonald et~al.}{2018}]{McDonald2018}
\begin{barticle}
\bauthor{\bsnm{McDonald}, \binits{A.}},
\bauthor{\bsnm{Pereg-Barnea}, \binits{T.}},
\bauthor{\bsnm{Clerk}, \binits{A.}}:
\batitle{Phase-dependent chiral transport and effective non-hermitian dynamics
  in a bosonic kitaev-majorana chain}.
\bjtitle{Phys. Rev. X}
\bvolume{8}(\bissue{4}),
\bfpage{041031}
(\byear{2018})
\doiurl{10.1103/physrevx.8.041031}
\end{barticle}
\endbibitem

\bibitem[\protect\citeauthoryear{Hatano and Nelson}{1996}]{HNmodel1}
\begin{barticle}
\bauthor{\bsnm{Hatano}, \binits{N.}},
\bauthor{\bsnm{Nelson}, \binits{D.R.}}:
\batitle{Localization transitions in non-hermitian quantum mechanics}.
\bjtitle{Phys. Rev. Lett.}
\bvolume{77},
\bfpage{570}--\blpage{573}
(\byear{1996})
\doiurl{10.1103/PhysRevLett.77.570}
\end{barticle}
\endbibitem

\bibitem[\protect\citeauthoryear{Hatano and Nelson}{1997}]{HNmodel2}
\begin{barticle}
\bauthor{\bsnm{Hatano}, \binits{N.}},
\bauthor{\bsnm{Nelson}, \binits{D.R.}}:
\batitle{Vortex pinning and non-hermitian quantum mechanics}.
\bjtitle{Phys. Rev. B}
\bvolume{56},
\bfpage{8651}--\blpage{8673}
(\byear{1997})
\doiurl{10.1103/PhysRevB.56.8651}
\end{barticle}
\endbibitem

\bibitem[\protect\citeauthoryear{Longhi et~al.}{2015a}]{Longhi2015}
\begin{barticle}
\bauthor{\bsnm{Longhi}, \binits{S.}},
\bauthor{\bsnm{Gatti}, \binits{D.}},
\bauthor{\bsnm{Valle}, \binits{G.D.}}:
\batitle{Robust light transport in non-hermitian photonic lattices}.
\bjtitle{Scientific reports}
\bvolume{5}(\bissue{1}),
\bfpage{13376}
(\byear{2015})
\doiurl{10.1038/srep13376}
\end{barticle}
\endbibitem

\bibitem[\protect\citeauthoryear{Longhi et~al.}{2015b}]{Longhi2015a}
\begin{barticle}
\bauthor{\bsnm{Longhi}, \binits{S.}},
\bauthor{\bsnm{Gatti}, \binits{D.}},
\bauthor{\bsnm{Della~Valle}, \binits{G.}}:
\batitle{Non-hermitian transparency and one-way transport in low-dimensional
  lattices by an imaginary gauge field}.
\bjtitle{Phys. Rev. B}
\bvolume{92}(\bissue{9}),
\bfpage{094204}
(\byear{2015})
\doiurl{10.1103/physrevb.92.094204}
\end{barticle}
\endbibitem

\bibitem[\protect\citeauthoryear{Martinez~Alvarez
  et~al.}{2018}]{MartinezAlvarez2018}
\begin{barticle}
\bauthor{\bsnm{Martinez~Alvarez}, \binits{V.}},
\bauthor{\bsnm{Barrios~Vargas}, \binits{J.}},
\bauthor{\bsnm{Foa~Torres}, \binits{L.}}:
\batitle{Non-hermitian robust edge states in one dimension: Anomalous
  localization and eigenspace condensation at exceptional points}.
\bjtitle{Phys. Rev. B}
\bvolume{97}(\bissue{12}),
\bfpage{121401}
(\byear{2018})
\doiurl{10.1103/physrevb.97.121401}
\end{barticle}
\endbibitem

\bibitem[\protect\citeauthoryear{Zhang et~al.}{2020}]{skinmode1}
\begin{barticle}
\bauthor{\bsnm{Zhang}, \binits{K.}},
\bauthor{\bsnm{Yang}, \binits{Z.}},
\bauthor{\bsnm{Fang}, \binits{C.}}:
\batitle{Correspondence between winding numbers and skin modes in non-hermitian
  systems}.
\bjtitle{Phys. Rev. Lett.}
\bvolume{125},
\bfpage{126402}
(\byear{2020})
\doiurl{10.1103/PhysRevLett.125.126402}
\end{barticle}
\endbibitem

\bibitem[\protect\citeauthoryear{Okuma et~al.}{2020}]{Topoorg}
\begin{barticle}
\bauthor{\bsnm{Okuma}, \binits{N.}},
\bauthor{\bsnm{Kawabata}, \binits{K.}},
\bauthor{\bsnm{Shiozaki}, \binits{K.}},
\bauthor{\bsnm{Sato}, \binits{M.}}:
\batitle{Topological origin of non-hermitian skin effects}.
\bjtitle{Phys. Rev. Lett.}
\bvolume{124},
\bfpage{086801}
(\byear{2020})
\doiurl{10.1103/PhysRevLett.124.086801}
\end{barticle}
\endbibitem

\bibitem[\protect\citeauthoryear{Longhi}{2019}]{Longhi2019}
\begin{barticle}
\bauthor{\bsnm{Longhi}, \binits{S.}}:
\batitle{Probing non-hermitian skin effect and non-bloch phase transitions}.
\bjtitle{Phys. Rev. Research}
\bvolume{1},
\bfpage{023013}
(\byear{2019})
\doiurl{10.1103/PhysRevResearch.1.023013}
\end{barticle}
\endbibitem

\bibitem[\protect\citeauthoryear{Song et~al.}{2019}]{song2019}
\begin{barticle}
\bauthor{\bsnm{Song}, \binits{F.}},
\bauthor{\bsnm{Yao}, \binits{S.}},
\bauthor{\bsnm{Wang}, \binits{Z.}}:
\batitle{Non-hermitian skin effect and chiral damping in open quantum systems}.
\bjtitle{Phys. Rev. Lett.}
\bvolume{123},
\bfpage{170401}
(\byear{2019})
\doiurl{10.1103/PhysRevLett.123.170401}
\end{barticle}
\endbibitem

\bibitem[\protect\citeauthoryear{Yokomizo and Murakami}{2021}]{Yokomizo2021}
\begin{barticle}
\bauthor{\bsnm{Yokomizo}, \binits{K.}},
\bauthor{\bsnm{Murakami}, \binits{S.}}:
\batitle{Scaling rule for the critical non-hermitian skin effect}.
\bjtitle{Phys. Rev. B}
\bvolume{104}(\bissue{16}),
\bfpage{165117}
(\byear{2021})
\doiurl{10.1103/physrevb.104.165117}
\end{barticle}
\endbibitem

\bibitem[\protect\citeauthoryear{Yang et~al.}{2020}]{Jiangping2020}
\begin{barticle}
\bauthor{\bsnm{Yang}, \binits{Z.}},
\bauthor{\bsnm{Zhang}, \binits{K.}},
\bauthor{\bsnm{Fang}, \binits{C.}},
\bauthor{\bsnm{Hu}, \binits{J.}}:
\batitle{Non-hermitian bulk-boundary correspondence and auxiliary generalized
  brillouin zone theory}.
\bjtitle{Phys. Rev. Lett.}
\bvolume{125},
\bfpage{226402}
(\byear{2020})
\doiurl{10.1103/PhysRevLett.125.226402}
\end{barticle}
\endbibitem

\bibitem[\protect\citeauthoryear{Xu et~al.}{2021}]{Xu2021a}
\begin{barticle}
\bauthor{\bsnm{Xu}, \binits{K.}},
\bauthor{\bsnm{Zhang}, \binits{X.}},
\bauthor{\bsnm{Luo}, \binits{K.}},
\bauthor{\bsnm{Yu}, \binits{R.}},
\bauthor{\bsnm{Li}, \binits{D.}},
\bauthor{\bsnm{Zhang}, \binits{H.}}:
\batitle{Coexistence of topological edge states and skin effects in the
  non-hermitian su-schrieffer-heeger model with long-range nonreciprocal
  hopping in topoelectric realizations}.
\bjtitle{Phys. Rev. B}
\bvolume{103}(\bissue{12}),
\bfpage{125411}
(\byear{2021})
\doiurl{10.1103/physrevb.103.125411}
\end{barticle}
\endbibitem

\bibitem[\protect\citeauthoryear{Claes and Hughes}{2021}]{Claes2021}
\begin{barticle}
\bauthor{\bsnm{Claes}, \binits{J.}},
\bauthor{\bsnm{Hughes}, \binits{T.L.}}:
\batitle{Skin effect and winding number in disordered non-hermitian systems}.
\bjtitle{Phys. Rev. B}
\bvolume{103}(\bissue{14}),
\bfpage{140201}
(\byear{2021})
\doiurl{10.1103/physrevb.103.l140201}
\end{barticle}
\endbibitem

\bibitem[\protect\citeauthoryear{Okuma and Sato}{2021}]{Okuma2021}
\begin{barticle}
\bauthor{\bsnm{Okuma}, \binits{N.}},
\bauthor{\bsnm{Sato}, \binits{M.}}:
\batitle{Non-hermitian skin effects in hermitian correlated or disordered
  systems: quantities sensitive or insensitive to boundary effects and
  pseudo-quantum-number}.
\bjtitle{Phys. Rev. Lett.}
\bvolume{126}(\bissue{17}),
\bfpage{176601}
(\byear{2021})
\doiurl{10.1103/physrevlett.126.176601}
\end{barticle}
\endbibitem

\bibitem[\protect\citeauthoryear{Yao et~al.}{2018}]{Yao_2018}
\begin{barticle}
\bauthor{\bsnm{Yao}, \binits{S.}},
\bauthor{\bsnm{Song}, \binits{F.}},
\bauthor{\bsnm{Wang}, \binits{Z.}}:
\batitle{Non-hermitian chern bands}.
\bjtitle{Phys. Rev. Lett.}
\bvolume{121},
\bfpage{136802}
(\byear{2018})
\doiurl{10.1103/PhysRevLett.121.136802}
\end{barticle}
\endbibitem

\bibitem[\protect\citeauthoryear{Fu et~al.}{2021}]{fu2020nonhermitian}
\begin{barticle}
\bauthor{\bsnm{Fu}, \binits{Y.}},
\bauthor{\bsnm{Hu}, \binits{J.}},
\bauthor{\bsnm{Wan}, \binits{S.}}:
\batitle{Non-hermitian second-order skin and topological modes}.
\bjtitle{Phys. Rev. B}
\bvolume{103},
\bfpage{045420}
(\byear{2021})
\doiurl{10.1103/PhysRevB.103.045420}
\end{barticle}
\endbibitem

\bibitem[\protect\citeauthoryear{Kim and Park}{2021}]{Kim2021}
\begin{barticle}
\bauthor{\bsnm{Kim}, \binits{K.-M.}},
\bauthor{\bsnm{Park}, \binits{M.J.}}:
\batitle{Disorder-driven phase transition in the second-order non-hermitian
  skin effect}.
\bjtitle{Physical Review B}
\bvolume{104}(\bissue{12}),
\bfpage{121101}
(\byear{2021})
\doiurl{10.1103/physrevb.104.l121101}
\end{barticle}
\endbibitem

\bibitem[\protect\citeauthoryear{Ghorashi et~al.}{2021}]{Ghorashi2021}
\begin{barticle}
\bauthor{\bsnm{Ghorashi}, \binits{S.A.A.}},
\bauthor{\bsnm{Li}, \binits{T.}},
\bauthor{\bsnm{Sato}, \binits{M.}},
\bauthor{\bsnm{Hughes}, \binits{T.L.}}:
\batitle{Non-hermitian higher-order dirac semimetals}.
\bjtitle{Phys. Rev. B}
\bvolume{104}(\bissue{16}),
\bfpage{161116}
(\byear{2021})
\doiurl{10.1103/physrevb.104.l161116}
\end{barticle}
\endbibitem

\bibitem[\protect\citeauthoryear{Okuma and Sato}{2019}]{Okuma2019}
\begin{barticle}
\bauthor{\bsnm{Okuma}, \binits{N.}},
\bauthor{\bsnm{Sato}, \binits{M.}}:
\batitle{Topological phase transition driven by infinitesimal instability:
  Majorana fermions in non-hermitian spintronics}.
\bjtitle{Phys. Rev. Lett.}
\bvolume{123},
\bfpage{097701}
(\byear{2019})
\doiurl{10.1103/PhysRevLett.123.097701}
\end{barticle}
\endbibitem

\bibitem[\protect\citeauthoryear{Xu et~al.}{2021}]{Xu2021}
\begin{barticle}
\bauthor{\bsnm{Xu}, \binits{X.}},
\bauthor{\bsnm{Xu}, \binits{H.}},
\bauthor{\bsnm{Mandal}, \binits{S.}},
\bauthor{\bsnm{Banerjee}, \binits{R.}},
\bauthor{\bsnm{Ghosh}, \binits{S.}},
\bauthor{\bsnm{Liew}, \binits{T.C.H.}}:
\batitle{Interaction-induced double-sided skin effect in an exciton-polariton
  system}.
\bjtitle{Phys. Rev. B}
\bvolume{103},
\bfpage{235306}
(\byear{2021})
\doiurl{10.1103/PhysRevB.103.235306}
\end{barticle}
\endbibitem

\bibitem[\protect\citeauthoryear{Lin et~al.}{2021}]{Lin_2021}
\begin{barticle}
\bauthor{\bsnm{Lin}, \binits{Z.}},
\bauthor{\bsnm{Ding}, \binits{L.}},
\bauthor{\bsnm{Ke}, \binits{S.}},
\bauthor{\bsnm{Li}, \binits{X.}}:
\batitle{Steering non-hermitian skin modes by synthetic gauge fields in optical
  ring resonators}.
\bjtitle{Opt. Lett.}
\bvolume{46}(\bissue{15}),
\bfpage{3512}
(\byear{2021})
\doiurl{10.1364/OL.431904}
\end{barticle}
\endbibitem

\bibitem[\protect\citeauthoryear{Sun et~al.}{2021}]{Sun2021}
\begin{barticle}
\bauthor{\bsnm{Sun}, \binits{X.-Q.}},
\bauthor{\bsnm{Zhu}, \binits{P.}},
\bauthor{\bsnm{Hughes}, \binits{T.L.}}:
\batitle{Geometric response and disclination-induced skin effects in
  non-hermitian systems}.
\bjtitle{Phys. Rev. Lett.}
\bvolume{127}(\bissue{6}),
\bfpage{066401}
(\byear{2021})
\doiurl{10.1103/physrevlett.127.066401}
\end{barticle}
\endbibitem

\bibitem[\protect\citeauthoryear{Yuce}{2021}]{Yuce2021}
\begin{barticle}
\bauthor{\bsnm{Yuce}, \binits{C.}}:
\batitle{Nonlinear non-hermitian skin effect}.
\bjtitle{Phys. Lett. A}
\bvolume{408},
\bfpage{127484}
(\byear{2021})
\doiurl{10.1016/j.physleta.2021.127484}
\end{barticle}
\endbibitem

\bibitem[\protect\citeauthoryear{Zhang et~al.}{2021a}]{zhang2021universal}
\begin{botherref}
\oauthor{\bsnm{Zhang}, \binits{K.}},
\oauthor{\bsnm{Yang}, \binits{Z.}},
\oauthor{\bsnm{Fang}, \binits{C.}}:
Universal non-Hermitian skin effect in two and higher dimensions
(2021)
\end{botherref}
\endbibitem

\bibitem[\protect\citeauthoryear{Zhang et~al.}{2021b}]{zhang2021acoustic}
\begin{botherref}
\oauthor{\bsnm{Zhang}, \binits{L.}},
\oauthor{\bsnm{Yang}, \binits{Y.}},
\oauthor{\bsnm{Ge}, \binits{Y.}},
\oauthor{\bsnm{Guan}, \binits{Y.-j.}},
\oauthor{\bsnm{Chen}, \binits{Q.}},
\oauthor{\bsnm{Yan}, \binits{Q.}},
\oauthor{\bsnm{Chen}, \binits{F.}},
\oauthor{\bsnm{Xi}, \binits{R.}},
\oauthor{\bsnm{Li}, \binits{Y.}},
\oauthor{\bsnm{Jia}, \binits{D.}},
\oauthor{\bsnm{Yuan}, \binits{S.-q.}},
\oauthor{\bsnm{Sun}, \binits{H.-x.}},
\oauthor{\bsnm{Chen}, \binits{H.}},
\oauthor{\bsnm{Zhang}, \binits{B.}}:
Acoustic non-Hermitian skin effect from twisted winding topology
(2021)
\end{botherref}
\endbibitem

\bibitem[\protect\citeauthoryear{Liang et~al.}{2022}]{Liang2022}
\begin{barticle}
\bauthor{\bsnm{Liang}, \binits{Q.}},
\bauthor{\bsnm{Xie}, \binits{D.}},
\bauthor{\bsnm{Dong}, \binits{Z.}},
\bauthor{\bsnm{Li}, \binits{H.}},
\bauthor{\bsnm{Li}, \binits{H.}},
\bauthor{\bsnm{Gadway}, \binits{B.}},
\bauthor{\bsnm{Yi}, \binits{W.}},
\bauthor{\bsnm{Yan}, \binits{B.}}:
\batitle{Dynamic signatures of non-hermitian skin effect and topology in
  ultracold atoms}.
\bjtitle{Phys. Rev. Lett.}
\bvolume{129}(\bissue{7}),
\bfpage{070401}
(\byear{2022})
\doiurl{10.1103/physrevlett.129.070401}
\end{barticle}
\endbibitem

\bibitem[\protect\citeauthoryear{Shang et~al.}{2022}]{Shang2022}
\begin{barticle}
\bauthor{\bsnm{Shang}, \binits{C.}},
\bauthor{\bsnm{Liu}, \binits{S.}},
\bauthor{\bsnm{Shao}, \binits{R.}},
\bauthor{\bsnm{Han}, \binits{P.}},
\bauthor{\bsnm{Zang}, \binits{X.}},
\bauthor{\bsnm{Zhang}, \binits{X.}},
\bauthor{\bsnm{Salama}, \binits{K.N.}},
\bauthor{\bsnm{Gao}, \binits{W.}},
\bauthor{\bsnm{Lee}, \binits{C.H.}},
\bauthor{\bsnm{Thomale}, \binits{R.}},
\bauthor{\bsnm{Manchon}, \binits{A.}},
\bauthor{\bsnm{Zhang}, \binits{S.}},
\bauthor{\bsnm{Cui}, \binits{T.J.}},
\bauthor{\bsnm{Schwingenschlögl}, \binits{U.}}:
\batitle{Experimental identification of the second‐order non‐hermitian skin
  effect with physics‐graph‐informed machine learning}.
\bjtitle{Adv. Sci.}
\bvolume{9}(\bissue{36}),
\bfpage{2202922}
(\byear{2022})
\doiurl{10.1002/advs.202202922}
\end{barticle}
\endbibitem

\bibitem[\protect\citeauthoryear{Alase et~al.}{2016}]{PhysRevLett.117.076804}
\begin{barticle}
\bauthor{\bsnm{Alase}, \binits{A.}},
\bauthor{\bsnm{Cobanera}, \binits{E.}},
\bauthor{\bsnm{Ortiz}, \binits{G.}},
\bauthor{\bsnm{Viola}, \binits{L.}}:
\batitle{Exact solution of quadratic fermionic hamiltonians for arbitrary
  boundary conditions}.
\bjtitle{Phys. Rev. Lett.}
\bvolume{117},
\bfpage{076804}
(\byear{2016})
\doiurl{10.1103/PhysRevLett.117.076804}
\end{barticle}
\endbibitem

\bibitem[\protect\citeauthoryear{Helbig et~al.}{2020}]{Helbig_2020}
\begin{barticle}
\bauthor{\bsnm{Helbig}, \binits{T.}},
\bauthor{\bsnm{Hofmann}, \binits{T.}},
\bauthor{\bsnm{Imhof}, \binits{S.}},
\bauthor{\bsnm{Abdelghany}, \binits{M.}},
\bauthor{\bsnm{Kiessling}, \binits{T.}},
\bauthor{\bsnm{Molenkamp}, \binits{L.W.}},
\bauthor{\bsnm{Lee}, \binits{C.H.}},
\bauthor{\bsnm{Szameit}, \binits{A.}},
\bauthor{\bsnm{Greiter}, \binits{M.}},
\bauthor{\bsnm{Thomale}, \binits{R.}}:
\batitle{Generalized bulk{\textendash}boundary correspondence in non-hermitian
  topolectrical circuits}.
\bjtitle{Nat. Phys.}
\bvolume{16}(\bissue{7}),
\bfpage{747}--\blpage{750}
(\byear{2020})
\doiurl{10.1038/s41567-020-0922-9}
\end{barticle}
\endbibitem

\bibitem[\protect\citeauthoryear{Liu et~al.}{2020}]{Liu2020}
\begin{barticle}
\bauthor{\bsnm{Liu}, \binits{C.-H.}},
\bauthor{\bsnm{Zhang}, \binits{K.}},
\bauthor{\bsnm{Yang}, \binits{Z.}},
\bauthor{\bsnm{Chen}, \binits{S.}}:
\batitle{Helical damping and dynamical critical skin effect in open quantum
  systems}.
\bjtitle{Phys. Rev. Research}
\bvolume{2},
\bfpage{043167}
(\byear{2020})
\doiurl{10.1103/PhysRevResearch.2.043167}
\end{barticle}
\endbibitem

\bibitem[\protect\citeauthoryear{Lee et~al.}{2019}]{Lee2019}
\begin{barticle}
\bauthor{\bsnm{Lee}, \binits{C.H.}},
\bauthor{\bsnm{Li}, \binits{L.}},
\bauthor{\bsnm{Gong}, \binits{J.}}:
\batitle{Hybrid higher-order skin-topological modes in nonreciprocal systems}.
\bjtitle{Phys. Rev. Lett.}
\bvolume{123},
\bfpage{016805}
(\byear{2019})
\doiurl{10.1103/PhysRevLett.123.016805}
\end{barticle}
\endbibitem

\bibitem[\protect\citeauthoryear{Xiao et~al.}{2020}]{Xiao_2020}
\begin{barticle}
\bauthor{\bsnm{Xiao}, \binits{L.}},
\bauthor{\bsnm{Deng}, \binits{T.}},
\bauthor{\bsnm{Wang}, \binits{K.}},
\bauthor{\bsnm{Zhu}, \binits{G.}},
\bauthor{\bsnm{Wang}, \binits{Z.}},
\bauthor{\bsnm{Yi}, \binits{W.}},
\bauthor{\bsnm{Xue}, \binits{P.}}:
\batitle{Non-hermitian bulk{\textendash}boundary correspondence in quantum
  dynamics}.
\bjtitle{Nat. Phys.}
\bvolume{16}(\bissue{7}),
\bfpage{761}--\blpage{766}
(\byear{2020})
\doiurl{10.1038/s41567-020-0836-6}
\end{barticle}
\endbibitem

\bibitem[\protect\citeauthoryear{Imura and Takane}{2019}]{Imura2019}
\begin{barticle}
\bauthor{\bsnm{Imura}, \binits{K.-I.}},
\bauthor{\bsnm{Takane}, \binits{Y.}}:
\batitle{Generalized bulk-edge correspondence for non-hermitian topological
  systems}.
\bjtitle{Phys. Rev. B}
\bvolume{100},
\bfpage{165430}
(\byear{2019})
\doiurl{10.1103/PhysRevB.100.165430}
\end{barticle}
\endbibitem

\bibitem[\protect\citeauthoryear{Xu et~al.}{2020}]{Xu_2020nh}
\begin{barticle}
\bauthor{\bsnm{Xu}, \binits{X.}},
\bauthor{\bsnm{Liu}, \binits{H.}},
\bauthor{\bsnm{Zhang}, \binits{Z.}},
\bauthor{\bsnm{Liang}, \binits{Z.}}:
\batitle{The non-hermitian geometrical property of 1d lieb lattice under
  majorana's stellar representation}.
\bjtitle{J. Phys-Condens. Mat.}
\bvolume{32}(\bissue{42}),
\bfpage{425402}
(\byear{2020})
\doiurl{10.1088/1361-648x/ab9fd4}
\end{barticle}
\endbibitem

\bibitem[\protect\citeauthoryear{Gong et~al.}{2018}]{Masahito2018}
\begin{barticle}
\bauthor{\bsnm{Gong}, \binits{Z.}},
\bauthor{\bsnm{Ashida}, \binits{Y.}},
\bauthor{\bsnm{Kawabata}, \binits{K.}},
\bauthor{\bsnm{Takasan}, \binits{K.}},
\bauthor{\bsnm{Higashikawa}, \binits{S.}},
\bauthor{\bsnm{Ueda}, \binits{M.}}:
\batitle{Topological phases of non-hermitian systems}.
\bjtitle{Phys. Rev. X}
\bvolume{8},
\bfpage{031079}
(\byear{2018})
\doiurl{10.1103/PhysRevX.8.031079}
\end{barticle}
\endbibitem

\bibitem[\protect\citeauthoryear{Lieu}{2018}]{Lieu2018}
\begin{barticle}
\bauthor{\bsnm{Lieu}, \binits{S.}}:
\batitle{Topological phases in the non-hermitian su-schrieffer-heeger model}.
\bjtitle{Phys. Rev. B}
\bvolume{97},
\bfpage{045106}
(\byear{2018})
\doiurl{10.1103/PhysRevB.97.045106}
\end{barticle}
\endbibitem

\bibitem[\protect\citeauthoryear{Xu et~al.}{2022}]{XU2022}
\begin{barticle}
\bauthor{\bsnm{Xu}, \binits{X.}},
\bauthor{\bsnm{Bao}, \binits{R.}},
\bauthor{\bsnm{Liew}, \binits{T.C.H.}}:
\batitle{Non-hermitian topological exciton-polariton corner modes}.
\bjtitle{Phys. Rev. B}
\bvolume{106},
\bfpage{201302}
(\byear{2022})
\doiurl{10.1103/PhysRevB.106.L201302}
\end{barticle}
\endbibitem

\bibitem[\protect\citeauthoryear{Longhi}{2022}]{PhysRevB.105.245143}
\begin{barticle}
\bauthor{\bsnm{Longhi}, \binits{S.}}:
\batitle{Non-hermitian skin effect and self-acceleration}.
\bjtitle{Phys. Rev. B}
\bvolume{105},
\bfpage{245143}
(\byear{2022})
\doiurl{10.1103/PhysRevB.105.245143}
\end{barticle}
\endbibitem

\bibitem[\protect\citeauthoryear{Weidemann et~al.}{2022}]{Weidemann_2022}
\begin{barticle}
\bauthor{\bsnm{Weidemann}, \binits{S.}},
\bauthor{\bsnm{Kremer}, \binits{M.}},
\bauthor{\bsnm{Longhi}, \binits{S.}},
\bauthor{\bsnm{Szameit}, \binits{A.}}:
\batitle{Topological triple phase transition in non-hermitian floquet
  quasicrystals}.
\bjtitle{Nature}
\bvolume{601}(\bissue{7893}),
\bfpage{354}--\blpage{359}
(\byear{2022})
\doiurl{10.1038/s41586-021-04253-0}
\end{barticle}
\endbibitem

\bibitem[\protect\citeauthoryear{Lin et~al.}{2022}]{PhysRevLett.129.113601}
\begin{barticle}
\bauthor{\bsnm{Lin}, \binits{Q.}},
\bauthor{\bsnm{Li}, \binits{T.}},
\bauthor{\bsnm{Xiao}, \binits{L.}},
\bauthor{\bsnm{Wang}, \binits{K.}},
\bauthor{\bsnm{Yi}, \binits{W.}},
\bauthor{\bsnm{Xue}, \binits{P.}}:
\batitle{Topological phase transitions and mobility edges in non-hermitian
  quasicrystals}.
\bjtitle{Phys. Rev. Lett.}
\bvolume{129},
\bfpage{113601}
(\byear{2022})
\doiurl{10.1103/PhysRevLett.129.113601}
\end{barticle}
\endbibitem

\bibitem[\protect\citeauthoryear{Parto et~al.}{2023}]{Parto_2023}
\begin{botherref}
\oauthor{\bsnm{Parto}, \binits{M.}},
\oauthor{\bsnm{Leefmans}, \binits{C.}},
\oauthor{\bsnm{Williams}, \binits{J.}},
\oauthor{\bsnm{Nori}, \binits{F.}},
\oauthor{\bsnm{Marandi}, \binits{A.}}:
Non-abelian effects in dissipative photonic topological lattices.
Nature Communications
\textbf{14}(1)
(2023)
\doiurl{10.1038/s41467-023-37065-z}
\end{botherref}
\endbibitem

\bibitem[\protect\citeauthoryear{Zhang et~al.}{2022}]{Zhang2022review}
\begin{botherref}
\oauthor{\bsnm{Zhang}, \binits{X.}},
\oauthor{\bsnm{Zhang}, \binits{T.}},
\oauthor{\bsnm{Lu}, \binits{M.-H.}},
\oauthor{\bsnm{Chen}, \binits{Y.-F.}}:
A review on non-hermitian skin effect.
Advances in Physics: X
\textbf{7}(1)
(2022)
\doiurl{10.1080/23746149.2022.2109431}
\end{botherref}
\endbibitem

\bibitem[\protect\citeauthoryear{Wang et~al.}{2023}]{scaling2023}
\begin{barticle}
\bauthor{\bsnm{Wang}, \binits{Y.-C.}},
\bauthor{\bsnm{Jen}, \binits{H.H.}},
\bauthor{\bsnm{You}, \binits{J.-S.}}:
\batitle{Scaling laws for non-hermitian skin effect with long-range couplings}.
\bjtitle{Phys. Rev. B}
\bvolume{108},
\bfpage{085418}
(\byear{2023})
\doiurl{10.1103/PhysRevB.108.085418}
\end{barticle}
\endbibitem

\bibitem[\protect\citeauthoryear{Qin et~al.}{2023}]{scalingfang}
\begin{barticle}
\bauthor{\bsnm{Qin}, \binits{F.}},
\bauthor{\bsnm{Ma}, \binits{Y.}},
\bauthor{\bsnm{Shen}, \binits{R.}},
\bauthor{\bsnm{Lee}, \binits{C.H.}}:
\batitle{Universal competitive spectral scaling from the critical non-hermitian
  skin effect}.
\bjtitle{Phys. Rev. B}
\bvolume{107},
\bfpage{155430}
(\byear{2023})
\doiurl{10.1103/PhysRevB.107.155430}
\end{barticle}
\endbibitem

\bibitem[\protect\citeauthoryear{Hu et~al.}{2023}]{PhysRevB.108.115404}
\begin{barticle}
\bauthor{\bsnm{Hu}, \binits{Y.-M.R.}},
\bauthor{\bsnm{Ostrovskaya}, \binits{E.A.}},
\bauthor{\bsnm{Estrecho}, \binits{E.}}:
\batitle{Wave-packet dynamics in a non-hermitian exciton-polariton system}.
\bjtitle{Phys. Rev. B}
\bvolume{108},
\bfpage{115404}
(\byear{2023})
\doiurl{10.1103/PhysRevB.108.115404}
\end{barticle}
\endbibitem

\bibitem[\protect\citeauthoryear{Malzard et~al.}{2018}]{Malzard_2018}
\begin{barticle}
\bauthor{\bsnm{Malzard}, \binits{S.}},
\bauthor{\bsnm{Cancellieri}, \binits{E.}},
\bauthor{\bsnm{Schomerus}, \binits{H.}}:
\batitle{Topological dynamics and excitations in lasers and condensates with
  saturable gain or loss}.
\bjtitle{Opt. Express}
\bvolume{26}(\bissue{17}),
\bfpage{22506}
(\byear{2018})
\doiurl{10.1364/OE.26.022506}
\end{barticle}
\endbibitem

\bibitem[\protect\citeauthoryear{Liu et~al.}{2020}]{Liu_2020}
\begin{barticle}
\bauthor{\bsnm{Liu}, \binits{J.S.}},
\bauthor{\bsnm{Han}, \binits{Y.Z.}},
\bauthor{\bsnm{Liu}, \binits{C.S.}}:
\batitle{A new way to construct topological invariants of non-hermitian systems
  with the non-hermitian skin effect}.
\bjtitle{Chin. Phys. B}
\bvolume{29}(\bissue{1}),
\bfpage{010302}
(\byear{2020})
\doiurl{10.1088/1674-1056/ab5937}
\end{barticle}
\endbibitem

\bibitem[\protect\citeauthoryear{Fu et~al.}{2020}]{Fu_2020}
\begin{barticle}
\bauthor{\bsnm{Fu}, \binits{Z.}},
\bauthor{\bsnm{Fu}, \binits{N.}},
\bauthor{\bsnm{Zhang}, \binits{H.}},
\bauthor{\bsnm{Wang}, \binits{Z.}},
\bauthor{\bsnm{Zhao}, \binits{D.}},
\bauthor{\bsnm{Ke}, \binits{S.}}:
\batitle{Extended {SSH} model in non-hermitian waveguides with alternating real
  and imaginary couplings}.
\bjtitle{Appl. Sci.}
\bvolume{10}(\bissue{10}),
\bfpage{3425}
(\byear{2020})
\doiurl{10.3390/app10103425}
\end{barticle}
\endbibitem

\bibitem[\protect\citeauthoryear{Yuan et~al.}{2023}]{Yuan_2023}
\begin{botherref}
\oauthor{\bsnm{Yuan}, \binits{H.}},
\oauthor{\bsnm{Zhang}, \binits{W.}},
\oauthor{\bsnm{Zhou}, \binits{Z.}},
\oauthor{\bsnm{Wang}, \binits{W.}},
\oauthor{\bsnm{Pan}, \binits{N.}},
\oauthor{\bsnm{Feng}, \binits{Y.}},
\oauthor{\bsnm{Sun}, \binits{H.}},
\oauthor{\bsnm{Zhang}, \binits{X.}}:
Non‐hermitian topolectrical circuit sensor with high sensitivity.
Advanced Science
\textbf{10}(19)
(2023)
\doiurl{10.1002/advs.202301128}
\end{botherref}
\endbibitem

\bibitem[\protect\citeauthoryear{Zhang et~al.}{2024}]{Zhang_2024}
\begin{botherref}
\oauthor{\bsnm{Zhang}, \binits{X.}},
\oauthor{\bsnm{Wu}, \binits{C.}},
\oauthor{\bsnm{Yan}, \binits{M.}},
\oauthor{\bsnm{Liu}, \binits{N.}},
\oauthor{\bsnm{Wang}, \binits{Z.}},
\oauthor{\bsnm{Chen}, \binits{G.}}:
Observation of continuum landau modes in non-hermitian electric circuits.
Nature Communications
\textbf{15}(1)
(2024)
\doiurl{10.1038/s41467-024-46122-0}
\end{botherref}
\endbibitem

\bibitem[\protect\citeauthoryear{Yu et~al.}{2024}]{Yu_2024}
\begin{barticle}
\bauthor{\bsnm{Yu}, \binits{T.}},
\bauthor{\bsnm{Zou}, \binits{J.}},
\bauthor{\bsnm{Zeng}, \binits{B.}},
\bauthor{\bsnm{Rao}, \binits{J.W.}},
\bauthor{\bsnm{Xia}, \binits{K.}}:
\batitle{Non-hermitian topological magnonics}.
\bjtitle{Physics Reports}
\bvolume{1062},
\bfpage{1}--\blpage{86}
(\byear{2024})
\doiurl{10.1016/j.physrep.2024.01.006}
\end{barticle}
\endbibitem

\bibitem[\protect\citeauthoryear{Yan et~al.}{2023}]{Yan_2023}
\begin{barticle}
\bauthor{\bsnm{Yan}, \binits{Q.}},
\bauthor{\bsnm{Zhao}, \binits{B.}},
\bauthor{\bsnm{Zhou}, \binits{R.}},
\bauthor{\bsnm{Ma}, \binits{R.}},
\bauthor{\bsnm{Lyu}, \binits{Q.}},
\bauthor{\bsnm{Chu}, \binits{S.}},
\bauthor{\bsnm{Hu}, \binits{X.}},
\bauthor{\bsnm{Gong}, \binits{Q.}}:
\batitle{Advances and applications on non-hermitian topological photonics}.
\bjtitle{Nanophotonics}
\bvolume{12}(\bissue{13}),
\bfpage{2247}--\blpage{2271}
(\byear{2023})
\doiurl{10.1515/nanoph-2022-0775}
\end{barticle}
\endbibitem

\bibitem[\protect\citeauthoryear{Li et~al.}{2023}]{Li_2023nn}
\begin{barticle}
\bauthor{\bsnm{Li}, \binits{A.}},
\bauthor{\bsnm{Wei}, \binits{H.}},
\bauthor{\bsnm{Cotrufo}, \binits{M.}},
\bauthor{\bsnm{Chen}, \binits{W.}},
\bauthor{\bsnm{Mann}, \binits{S.}},
\bauthor{\bsnm{Ni}, \binits{X.}},
\bauthor{\bsnm{Xu}, \binits{B.}},
\bauthor{\bsnm{Chen}, \binits{J.}},
\bauthor{\bsnm{Wang}, \binits{J.}},
\bauthor{\bsnm{Fan}, \binits{S.}},
\bauthor{\bsnm{Qiu}, \binits{C.-W.}},
\bauthor{\bsnm{Alù}, \binits{A.}},
\bauthor{\bsnm{Chen}, \binits{L.}}:
\batitle{Exceptional points and non-hermitian photonics at the nanoscale}.
\bjtitle{Nature Nanotechnology}
\bvolume{18}(\bissue{7}),
\bfpage{706}--\blpage{720}
(\byear{2023})
\doiurl{10.1038/s41565-023-01408-0}
\end{barticle}
\endbibitem

\bibitem[\protect\citeauthoryear{Zhao et~al.}{2019}]{Zhao_2019}
\begin{barticle}
\bauthor{\bsnm{Zhao}, \binits{H.}},
\bauthor{\bsnm{Qiao}, \binits{X.}},
\bauthor{\bsnm{Wu}, \binits{T.}},
\bauthor{\bsnm{Midya}, \binits{B.}},
\bauthor{\bsnm{Longhi}, \binits{S.}},
\bauthor{\bsnm{Feng}, \binits{L.}}:
\batitle{Non-hermitian topological light steering}.
\bjtitle{Science}
\bvolume{365}(\bissue{6458}),
\bfpage{1163}--\blpage{1166}
(\byear{2019})
\doiurl{10.1126/science.aay1064}
\end{barticle}
\endbibitem

\bibitem[\protect\citeauthoryear{Gu et~al.}{2022}]{Gu_2022}
\begin{botherref}
\oauthor{\bsnm{Gu}, \binits{Z.}},
\oauthor{\bsnm{Gao}, \binits{H.}},
\oauthor{\bsnm{Xue}, \binits{H.}},
\oauthor{\bsnm{Li}, \binits{J.}},
\oauthor{\bsnm{Su}, \binits{Z.}},
\oauthor{\bsnm{Zhu}, \binits{J.}}:
Transient non-hermitian skin effect.
Nature Communications
\textbf{13}(1)
(2022)
\doiurl{10.1038/s41467-022-35448-2}
\end{botherref}
\endbibitem

\end{thebibliography}

\end{document}